\documentclass[aps,pre,twocolumn,showpacs]{revtex4}

\usepackage{graphicx}

\begin{document}

\title{Scaling Distributions of Quarks, Mesons and Proton for all $p_T$,
Energy and Centrality}

\author{Rudolph C. Hwa$^1$ and C.\ B.\ Yang$^{1,2}$}

\affiliation{$^1$Institute of Theoretical Science and Department of Physics\\
University of Oregon, Eugene, OR 97403-5203, USA\\
$^2$Institute of Particle Physics, Hua-Zhong Normal University, Wuhan
430079, China}

\date{February 2003}

\begin{abstract}

We present the evidences for the existence of a universal scaling
behavior of the production of $\pi^0$ at all transverse momenta in
heavy-ion collisions at all centralities and all collision
energies. The corresponding scaling behavior of the quarks is then
derived just before the quarks  recombine with antiquarks to form
the pions. The degradation effect of the dense medium on the quark
$p_T$  is derived from the scaling distribution. In the
recombination model it is then possible to calculate the $p_T$
distributions of the produced proton and kaon, which are scaling
also. Experimentally verifiable predictions are made. Implications
of the existence of the scaling behavior are discussed.

\end{abstract}

\maketitle

\section{Introduction}

In two recent papers we have discussed the scaling properties of
the large $p_T$ distributions of $\pi^0$ produced in Au+Au
collisions at the relativistic heavy-ion collider (RHIC) and
presented their implications. As reported in the first paper
\cite{1}, hereafter referred to as I, we found energy scaling at
maximum centrality, while in the second paper \cite{2}, referred
to as II, we found centrality scaling at the highest energy. The
two can be combined to yield one scaling distribution for all
energy  and centrality. In I we derive the quark distribution
from the $\pi^0$ data in the framework of the recombination model;
we now do the same  for all centralities and determine the nature
of degradation of the quark momentum in the dense medium, as we
have done for the $\pi^0$ momentum in II. From the quark
distribution we can calculate the proton distribution for all
centralities. Moreover, we can extend our consideration to the
production of kaons so that we can calculate not only the $p/\pi$
ratio, but also the $K/\pi$ ratio.

We are able to do all that for two essential reasons. The first is
that the discovery of the scaling properties facilitates the
analysis by avoiding the need to consider the variation of physics
issues at different centralities and energies. The other is that
the quark distribution we derive is for $q$ and $\bar q$ just
before hadronization. It is not the result of some dynamical
evolution starting from hard collisions, for which many complex
issues must be considered \cite{3,4,5}. The recombination model
that we use can only address the hadronization problem of the soft
partons at low virtuality, but at any $p_T$. From the pion data we
infer the distributions of the soft $q$ and $\bar q$, which in
turn are used to give the proton and kaon distributions through
recombination. How the soft partons get to be where they are in
the $p_T$ space is not considered. However, by studying the
momentum degradation of the quarks, we gain from the centrality
dependence of the $p_T$ distributions some understanding about how
quarks lose momenta as they propagate through the dense medium.
Experimentally verifiable predictions are made on the proton and
kaon transverse momentum distributions.

The physical interpretation of the scaling variable is given at
the very end, where the term transversity is suggested to refer to
the difficulty of acquiring transverse motion. The broader
implication on the creation of quark-gluon plasma is finally
addressed.

\section{Scaling Distribution of Pions}

From the preliminary PHENIX data of $\pi^0$ production in Au+Au
collisions at RHIC \cite{6} we have found a scaling distribution
at midrapidity
\begin{equation}
\Phi(z)=A(N)K^2(s,N){1\over 2\pi p_T}{d^2N_\pi\over d\eta dp_T}\ ,
       \label{1}
\end{equation}
where
\begin{equation}
z=p_T/K(s,N)\ .       \label{2}
\end{equation}
The symbol $N$ denotes the number of participants, $N_{\rm part}$, for
brevity. The scaling factor $K(s,N)$ is first found in I for $N$
fixed at its maximum, $N_{\rm max}=350$, to be (in GeV/c)
\begin{equation}
K(s)=0.69+1.55\times10^{-3}\sqrt{s}\ ,      \label{3}
\end{equation}
where $\sqrt s$ is in units of GeV, and $K(s)$ is normalized to 1
at $\sqrt s=200$ GeV. When $N$ is allowed to vary, while $\sqrt s$
is fixed at 200 GeV, it is found in II that the scaling factor is
\begin{equation}
K(N)=1.226-6.36\times10^{-4}N\ ,         \label{4}
\end{equation}
normalized to 1 at $N=350$. We now combine the two and assume the
factorizable form
\begin{equation}
K(s,N)=K(s)K(N)\ .         \label{5}
\end{equation}
In this paper we investigate the centrality dependence mostly at
$\sqrt s=200$ GeV.
The normalization factor $A(N)$ in Eq.\ (\ref{1}) is found in II
to have a power-law dependence on the number of binary
collisions, $N_c$,
\begin{equation}
A(N_c)=530N_c^{-0.9}\ ,        \label{6}
\end{equation}
where $N_c$ in turn depends on $N$ as
\begin{equation}
N_c=0.44N^{1.33}\ ,        \label{7}
\end{equation}
a relationship that is determined from the tables listed in Refs.\
\cite{7,8}.
\begin{figure}[tbph]
\includegraphics[width=0.45\textwidth]{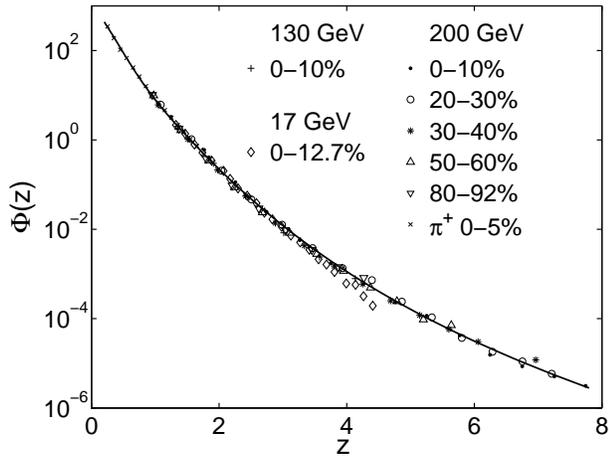}
\caption{Scaling distribution of pion production. Data at
$\sqrt s=130$ and 200 GeV are preliminary and were presented by
PHENIX \cite{6,9} for Au+Au collisions. The data at 17 GeV are
from WA98 \cite{8.1} for Pb+Pb collisions. The solid line is a fit
by Eq.\ (\ref{8}).}
\end{figure}

In Fig.\ 1 we show the combined plot of $\Phi(z)$ exhibiting the
$\sqrt s=200$ GeV PHENIX data for 5 bins of centrality as well as
the $\sqrt s=130$ GeV data at 0-10\% centrality \cite{6} and the 17
GeV data at 0-12.7\%  \cite{8.1}. The $\pi^0$ data, which are only
for $p_T>1$ GeV/c, are supplemented by the
$\pi^+$ data for $p_T<1$ GeV/c at
$\sqrt s=200$ GeV and 0-5\% centrality \cite{9}. Evidently, all the
data points fall on one universal scaling distribution that is
invariant under changes in $N$ and $\sqrt s$. The 17 GeV data are
obtained by the WA98 Collaboration \cite{8.1} for Pb+Pb collisions
and show a slight departure from the universal curve for $z>3$. It
should be recognized that those data points that deviate from the
scaling distribution correspond to
$p_T>3$ GeV/c, which is a $p_T$ range that represents a very large
fraction of the total available energy of 17 GeV. Thus the
kinematic constraint of energy conservation introduces a
non-dynamical factor that suppresses the high-$p_T$ behavior, not
present in the other data at $\sqrt s\ge$130 GeV. Such a
violation of scaling is expected, and should not be regarded as
an invalidation of the general scaling behavior we observe. On the
contrary, it is amazing that the scaling behavior can cover such a
wide range of $\sqrt s$, when most of the 17 GeV data points with
$p_T<3$ GeV/c are included.

The scaling data points can be well fitted by \cite{2}
\begin{equation}
\Phi(z)=1200\,(z^2+2)^{-4.8}\left(1+25\,{\rm e}^{-4.5z}\right)\ ,
\label{8}
\end{equation}
which is shown  by the solid line in Fig.\ 1. This formula differs
from the one given in I mainly by the addition of the
exponential term, which reflects the statistical behavior at small
$z$ represented by the $\pi^+$ data, but is insignificant at large
$z$, where the power-law behavior is indicative of the effects of
hard collisions.

A number of consequences of the scaling distribution, Eq.\
(\ref{8}), can be examined directly. First, the integral
\begin{equation}
{\cal I}=\int_0^\infty dz z\Phi(z)={A(N)\over 2\pi}{dN_{\pi^0}\over
d\eta}        \label{9}
\end{equation}
can be evaluated to yield ${\cal I}=46.2$. Using Eqs.\ (\ref{6}) and
(\ref{7}) for $A(N)=A(N_c(N))$, we can calculate the $N$
dependence of $dN_{\pi^0}/d\eta$ at midrapidity. The data that are
available for comparison are $dN_{\rm ch}/d\eta/(0.5N)$ vs $N$ at
$\sqrt s=130$ GeV \cite{8}, shown in Fig.\ 2. We plot in that
figure in solid line our calculated result for
$dN_{\pi^\pm}/d\eta/(0.5N)$, which is obtained from Eq.\ (\ref{9})
by identifying $dN_{\pi^\pm}/d\eta$ with $2dN_{\pi^0}/d\eta$.
Since $\pi^\pm$ is the dominant part of all charged particles, one
should regard the comparison between the calculated $N_{\pi^\pm}$
and the measured $N_{\rm ch}$ to be satisfactory. Recall that although the
$p/\pi$ ratio can exceed 1 around $p_T\approx 3$ GeV/c, it is small at
small $p_T$
where the distributions are dominant, so the production of proton does
not contribute to the integrated result as a large fraction of the total
$dN_{\rm ch}/d\eta$.
\begin{figure}[tbph]
\includegraphics[width=0.45\textwidth]{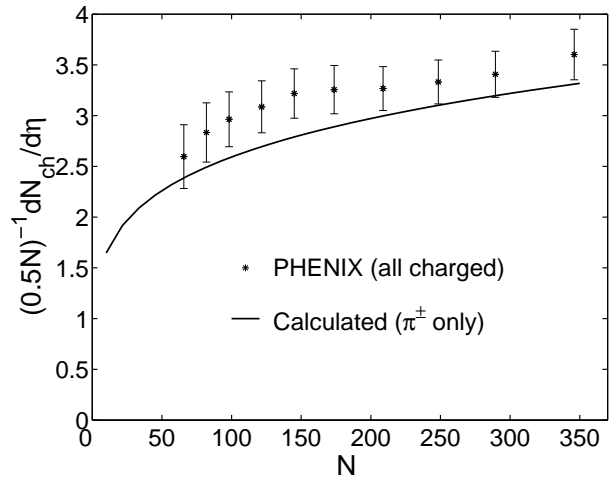}
\caption{Calculated $dN_{\pi^\pm}/d\eta/(0.5N)$ compared to
the data on $dN_{\rm ch}/d\eta/(0.5N)$ \cite{8}.}
\end{figure}

Another consequence of $\Phi(z)$ is the possibility to calculate
the nuclear modification factor
\begin{equation}
R_{AA}(p_T)={d^2N_{\pi^0}^{AA}/d\eta dp_T\over
N_c\,d^2N_{\pi^0}^{pp}/d\eta dp_T}\ .   \label{10}
\end{equation}
The data available for that are given in Ref.\ \cite{6}. We show
in Fig.\ 3 the data for two centrality bins: 0-10\% and 70-80\%.
The corresponding values of $N$ are 317 and 9.5, respectively
\cite{7}, for which the calculated $R_{AA}(p_T)$ are shown by the
solid and dash-dot lines. For $N_{\pi^0}^{pp}$ in the denominator
of Eq.\ (\ref{10}) we have used $N=2$. Experimentally, it is known
that the peripheral nuclear collisions
cannot be identified with $pp$ collisions. The ratio $R_{AA}(p_T)$
is defined with $N_{\pi^0}^{pp}$ in the denominator so as to have a
definitive experimental normalization. We calculate the ratio with
$N=2$ in the denominator so that no additional experimental input
is used. Our point is to show the consistency of $\Phi(z)$ when the
value of $N$ is extrapolated to extreme limits. Toward that end we
find the comparison
between the data points and the calculated curves in Fig.\ 3 to be
acceptable.
\begin{figure}[tbph]
\includegraphics[width=0.45\textwidth]{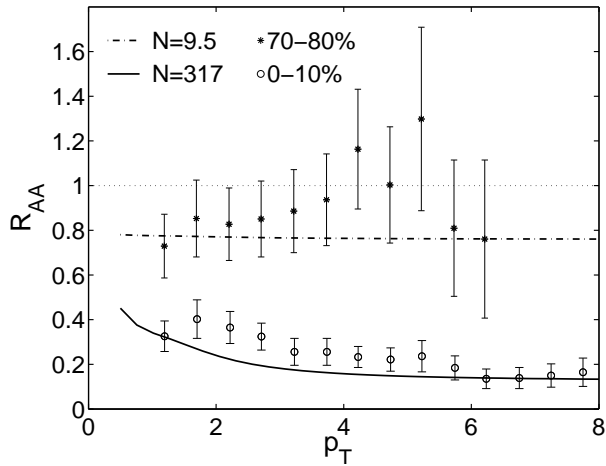}
\caption{Nuclear modification factor $R_{AA}(p_T)$ for
$\pi^0$ production. The preliminary data are from PHENIX \cite{6}.
The lines are calculated results using $N=2$ extrapolation of
$\Phi(z)$ for normalization rather than using independent $pp$
distribution.}
\end{figure}

The scaling distribution $\Phi(z)$ is a representation of the data
over 9 orders of magnitude for all centralities and for all
energies where data exist. Since the fit is done in the log scale
in Fig.\ 1, one can expect some deviations in the linear scale, as
the tests in Figs. 2 and 3 are done. The data must, of course, be
self-consistent, so any discrepencies in those figures are due to
the extrapolation of $\Phi(z)$ to very small $z$ in the
calculation of $dN_{\pi^0}/d\eta$ and to very small $N$ for $R_{AA}$.
We conclude from those tests that the formula (\ref{8}) for $\Phi(z)$
is quite reliable even down to very small values of $z$ and $N$.

\section{Scaling Distribution of Quarks}

From the scaling distribution of $\pi^0$ it is possible to derive
the quark distribution in the framework of the recombination
model. Those quarks have low virtuality (hence, soft) and are at
the last stage of their existence just before hadronization. They
are not to be confused with the  partons at high virtuality
(hence, hard) just after hard collisions. The evolution from the
hard partons to the soft quarks through gluon radiation and
conversion to quark pairs in the dense medium involve both
perturbative and non-perturbative QCD processes that are
complicated, only some of which can be calculated
\cite{3,5,10,11}. The last step to hadrons is circumvented by use
of the phenomenological fragmentation function that connects hard
partons to hadrons directly. Our use of the recombination model
treats only the last step from soft quarks to hadrons, and can
make no statement about the evolutionary process that begins from
hard collisions. What can be treated is how the soft quarks
recombine in different combinations to form different hadrons. That
is what we shall do in the following sections. Here we first
derive the quark distributions and examine how they depend on
centrality.

The application of the recombination model \cite{12,13} to the
high $p_T$ problem has been discussed in I. The recombination of
a $q$ and a $\bar q$ to form a $\pi^0$, where $q\bar q$ can be
either $u\bar u$ or $d\bar d$, but not both $u\bar u$ and $d\bar
d$, is described by [see I-Eq.\ (\ref{13})]
\begin{equation}
{dN_{\pi^0}\over zdz}=\int dz_1dz_2\,z_1z_2 F_{q\bar{q}}(z_1,z_2)
{R}_\pi(z_1,z_2,z)\ ,   \label{11}
\end{equation}
where our $dN_{\pi^0}/zdz$ is averaged over rapidity at
midrapidity and over the azimuthal angle $\phi$ with the $1/2\pi$
factor included, unlike the experimental distribution
$(2\pi)^{-1}d^2N_{\pi^0}/d\eta p_Tdp_T$, which is an average over
$\phi$  that shows the $1/2\pi$ factor explicitly, as in Eq.\
(\ref{1}). The joint distribution
$F_{q\bar q}(z_1,z_2)$ is assumed, for heavy-ion collisions, to
have the factorizable form
$F_{q\bar q}(z_1,z_2)=F_q(z_1)F_{\bar q}(z_2)$ with $F_q(z_1)$
being the quark distribution in the scaling variable $z_1$, and
$F_{\bar q}(z_2)$ for the antiquark. The recombination function
$R_\pi(z_1,z_2,z)$ depends on the wave function of the constituent
quarks in the pion; in the valon model \cite{1,13} it is
\begin{equation}
{R}_\pi(z_1,z_2,z)=z^{-2}\,\delta\left({z_1\over
z}+{z_2\over z}-1\right)\ ,   \label{12}
\end{equation}
where the valon distribution in the pion is determined by use of
the data on Drell-Yan production by pion, which is the only way to
probe the pion structure.

At $N=N_{\rm max}$, considered in I, the LHS of Eq.\ (\ref{11})
is identified with $\Phi(z)$, which in I is scaling in $s$. Now for
all centrality the LHS of Eq.\ (\ref{11}) is replaced by the new
$\Phi(z)$ given in Eq.\ (\ref{8}), scaling in both $N$ and $s$, and
on the RHS the new scaling $F_q(z_1)$ and
$F_{\bar q}(z_2)$ are to be determined. Thus putting the various
pieces together, we have
\begin{equation}
\Phi(z)=\int_0^z dz_1
z_1\left(1-\frac{z_1}{z}\right)F_q(z_1)F_{\bar{q}}(z-z_1)\ .
  \label{13}
\end{equation}
There is no explicit dependence on $N$ or $s$ in this equation,
but the $q$ and $\bar q$ distributions, being scale invariant,
have implicit dependences on $p_T$ and $N$ that will be examined
below.

Since the $q$ and $\bar q$ in Eq.\ (\ref{13}) are at the end of
their evolutionary processes, and are therefore soft partons
dominated by the products of gluon conversion, their distributions
can differ in normalization, but not significantly in their $z_1$
and $z_2$ dependences, a property that is supported by the
observation that the $\bar p/p$ ratio is nearly constant in $p_T$
\cite{14}. Denoting the $\bar p/p$ ratio by $c$,
 we thus use the relationship
\begin{equation}
F_{\bar{q}}(z)=c^{1/3}F_q(z)\ .          \label{14}
\end{equation}
At RHIC energies $c$ is roughly 0.7; at SPS
$c$ is about 0.13 \cite{14.1,14.2}. Since $c$ depends on the energy
$\sqrt s$, Eq.\ (\ref{14}) is not strictly a scaling relationship.
The smallness of $c$ at SPS energy is due to the difficulty of
producing a large number of $\bar p$ compared to
$p$ at $\sqrt s=17$ GeV \cite{14.3}. Such a scaling violation is
expected, just like the deviation of the diamond points at $z>3$
from the universal scaling curve in Fig.\ 1 for kinematical reasons.
However, over the whole range of variation of
$F_q(z)$ that we shall determine, the effect due to the variation of
$c$ is small by comparison, as we shall see.
\begin{figure}[tbph]
\includegraphics[width=0.45\textwidth]{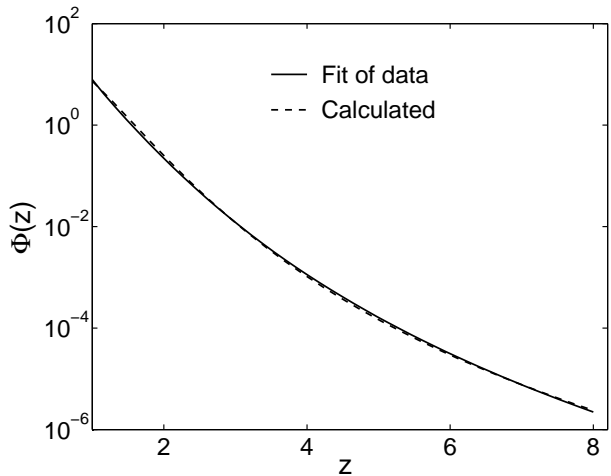}
\caption{Comparison of fitted and calculated curves of
$\Phi(z)$. The solid line is a plot of Eq.\ (\ref{8}) obtained
from the fit in Fig.\ 1. The dashed line is the result of the
calculation using Eq.\ (\ref{13}).}
\end{figure}

Let us put our emphasis on the scaling region of $c$ by setting
$c=0.7$. Using that in
Eq.\ (\ref{14}) and then on the RHS of (\ref{13}), with Eq.\
(\ref{8}) on the LHS, we can vary the parametrization of $F_q(z)$
to achieve a good fit of $\Phi(z)$. Our result is
\begin{equation}
F_q(z)=15\,(z^2+0.47z+0.72)^{-4.25}\ .           \label{15}
\end{equation}
In Fig.\ 4 we show a plot of Eq.\ (\ref{8}) by the solid line and
our fit of it using Eq.\ (\ref{15}) in (\ref{13}) by the dashed
line. Clearly, the fit is very good. The quark distribution given
by Eq.\ (\ref{15}) is shown in Fig.\ 5 over a range of 9 orders of
magnitude of variation. How would that be affected, if $c$ is
lowered to 0.13? The normalization of $F_q(z)$ in Fig.\ 5 would be
increased by a factor of $(0.7/0.13)^{1/6}=1.3$, which is
insignificant compared to the 9 orders of magnitude of variation of
$F_q(z)$.
\begin{figure}[tbph]
\includegraphics[width=0.45\textwidth]{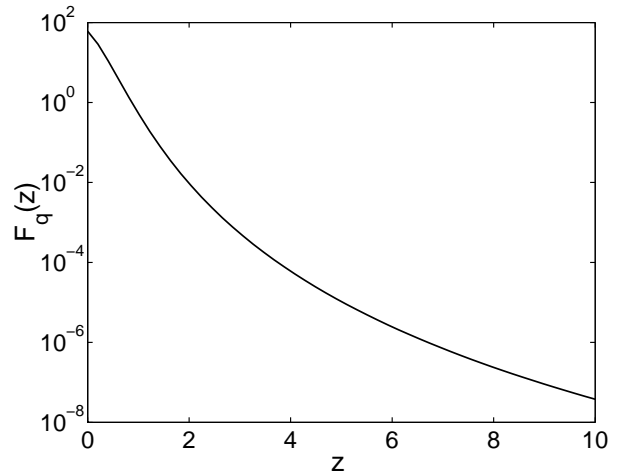}
\caption{Scaling distribution of quarks before hadronization.}
\end{figure}

From Eq.\ (\ref{15}) we can first determine the average $z$
\begin{equation}
\langle z\rangle={\int dz z^2 F_q(z)\over \int dz z F_q(z)}=0.415 .
   \label{16}
\end{equation}
We then define a new variable
\begin{equation}
u=z/\langle z\rangle\ ,      \label{17}
\end{equation}
in terms of which we can define a new distribution
\begin{eqnarray}
\psi_q(u) &=& F_q(z(u))\left/\int du\,u F_q(z(u))\right. \\
\nonumber
 &=&\langle z\rangle^2 F_q(z(u))\left/\int dz\,z F_q(z)\right.\ .
\label{18}
\end{eqnarray}
This distribution has not only the property that
\begin{equation}
\int du\ u\,\psi_q(u)=1            \label{19}
\end{equation}
by definition, but also
\begin{equation}
\int du\ u^2\,\psi_q(u)=1\ .           \label{20}
\end{equation}
These are the properties of a KNO-type distribution \cite{15}. To
be strictly KNO scaling, all higher moments should be independent
of $N$, as we shall investigate. The virtue of the scaling
variable $u$ is that it can be expressed directly in terms of
$p_T$, since from Eqs.\ (\ref{2}) and (\ref{17}) we have
\begin{equation}
u=p_T/\langle p_T\rangle_N\ ,               \label{21}
\end{equation}
where $\left<p_T\right>_N$ is the average $p_T$ at fixed
centrality and $s$. Note that the scale factor $K(s,N)$ is common
for the $p_T$ of both $\pi^0$ and  $q$. In the case of pions \cite{2}
the variable
$u$ can, in principle, be determined unambiguously from the
experiments directly, unlike the variable $z$ that requires
rescaling and fitting of the data at each $N$ and $s$. Indeed, the
variable $u$ is constructed in the same spirit as in the original
derivation of KNO scaling \cite{15}. Although in the case of quarks
here $u$ cannot be experimentally measured, it is useful to have a KNO
distribution $\psi_q(u)$ as a goal for theoretical modeling, since the
$u$ variable can more directly be related to the $p_T$ of the quarks.
\begin{figure}[tbph]
\includegraphics[width=0.45\textwidth]{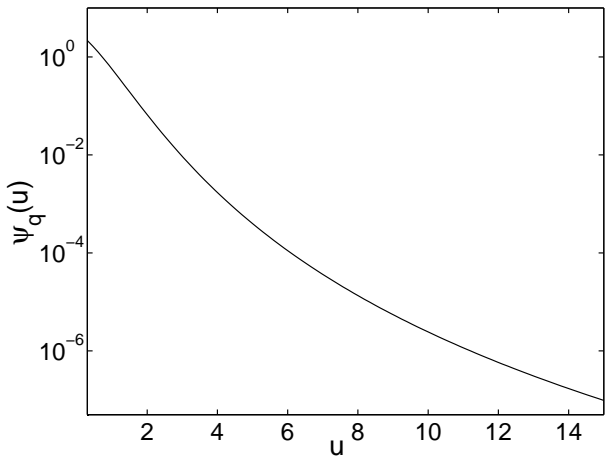}
\caption{ KNO-type distribution of quarks.}
\end{figure}
\begin{figure}[tbph]
\includegraphics[width=0.45\textwidth]{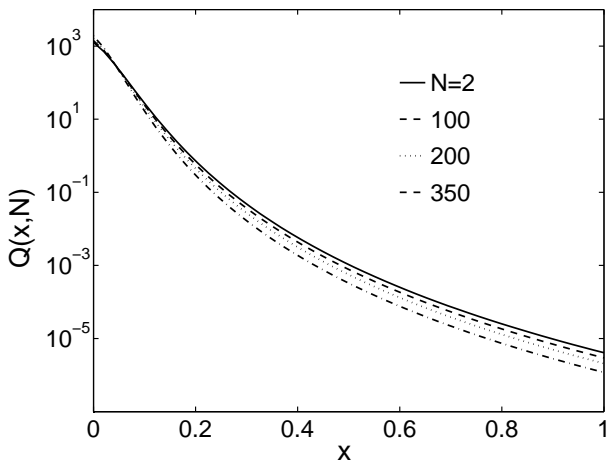}
\caption{Centrality dependence of normalized quark
distribution at $\sqrt s=200$ GeV.}
\end{figure}

From the definition in Eq.\ (18) we have
\begin{equation}
\psi_q(u)=1420\,(u^2+1.13u+4.18)^{-4.25}\ .     \label{22}
\end{equation}
A plot of this distribution is shown in Fig.\ 6. The existence of
such a scaling distribution in terms of an intuitive variable
given in Eq.\ (\ref{21}) suggests that there is a great deal of
regularity in the interplay between $N$ and $p_T$. Remembering how
$J/\psi$ production is expected to have an anomalous $p_T$
dependence in $AA$ collisions when deconfinement occurs \cite{16},
we find the lack of irregularity in the $p_T$ distribution of the
quarks to suggest that  the subject of $p_T$ dependences of
low-mass particles is not a fertile ground to find signals of
quark-gluon plasma --- unless, of course, scaling violation is
dramatically found at LHC, say.

Having analytic forms  for $F_q(z)$ and $\psi_q(u)$ enables us to
investigate the degradation of the parton $p_T$ in the dense
medium. To that end we define the momentum fraction variable
\begin{equation}
x=p_T/K_0\ ,               \label{23}
\end{equation}
where $K_0$ is a fixed scale, which we take to be 10 Gev/c. It is
tacitly assumed that there is no physics of interest here for
$p_T>10$ GeV/c. If that is not the case, an upward revision of
$K_0$ is trivial. From Eqs. (\ref{2}) and (\ref{23}) we have
\begin{equation}
z=x\,K_0/K(s,N)\ ,               \label{24}
\end{equation}
and we may rewrite the scaling quark distribution as
\begin{equation}
F_q(z)=F_q(x,N)\ ,               \label{25}
\end{equation}
where the $s$ dependence is suppressed. We can then define the
normalized quark distribution
\begin{equation}
Q(x,N)=F_q(x,N)\left/\int_0^1 dx\,x F_q(x, N)\right.\ .
\label{26}
\end{equation}
The $x$ dependences of $Q(x,N)$ for various representative values
of $N$ are shown in Fig.\ 7 for $\sqrt s=200$ GeV. It is evident that,
as $N$ increases, the high $x$ tail of $Q(x,N)$ is suppressed, with a
concurrent slight increase at very small $x$, since the integral of
$Q(x,N)$ over $xdx$ is constant at 1. The crossover is at $x\approx
0.04$, corresponding to $p_T\approx 0.4$ Gev/c. Thus the effect of the
dense medium is to degrade the $p_T$ of the quarks, which is a well
known property, but in a very regular way that we now  describe.

Let us define the ratio
\begin{equation}
R(x,N)={Q(x,N)\over Q(x,2)}           \label{27}
\end{equation}
so that the suppression at most $x$ and enhancement at small $x$
can be exhibited more clearly, as shown in Fig.\ 8 for $\sqrt s=200$
GeV. Note that for
$x>0.3$ the suppression is rather uniform. The rapid change in the
range $0<x<0.3$ can be seen in a different plot, shown in Fig.\ 9.
The decrease for $x>0.04$, as $N$ is increased, is now clear, as
is the increase for $x<0.04$. It should be recognized that in
normalizing $Q(x,N)$ by $Q(x,2)$ in Eq.\ (\ref{27}) it is not
important whether $Q(x,2)$ agrees well with the corresponding
distribution in $pp$ collisions. The ratio removes the exponential
dependence at small $x$ that is common for all $N$ and displays
better the relative change as $N$ is varied. Also, the
extrapolation to very small $x$ may not be accurate. The line for
$x=0.01$ in Fig.\ 9 is intended mainly to give a rough idea of the
nature of increase for $x<0.04$.
\begin{figure}[tbph]
\includegraphics[width=0.45\textwidth]{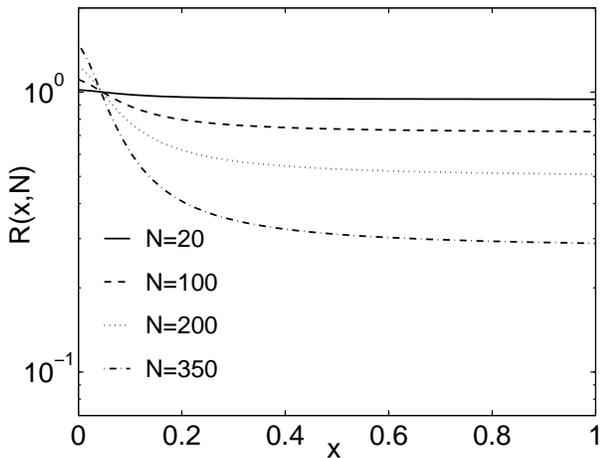}
\caption{Ratio of quark distribution relative to $N=2$,
exhibiting the degradation at $x>0.04$ and the enhancement at
$x<0.04$ for $\sqrt s=200$ GeV.}
\end{figure}

\begin{figure}[tbph]
\includegraphics[width=0.45\textwidth]{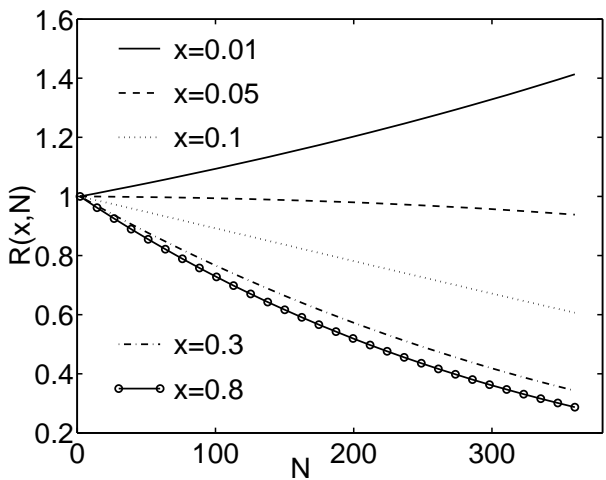}
\caption{Same as Fig.\ 8 but plotted for various fixed
values of $x$.}
\end{figure}

To quantify the degree of degradation, we take the moments of
$Q(x,N)$ and define
\begin{equation}
Q_n(N)=\int_0^1 dx\,x^{n+1}Q(x, N)\ .            \label{28}
\end{equation}
We do not consider the order of the moment $n>5$, since higher
moments demand more accuracy at higher $x$, which we cut off at
$x=1$. The extrapolation of our scaling distribution $F_q(z)$ to
higher $z$ is not without uncertainties. However, for $n\le 5$ our
analysis is reliable, and provides adequate insight into the
nature of the degradation in the $0<x<1$ region. In Fig.\ 10 we
show ln\,$Q_n(N)$ vs $N$ for $n=1,\cdots,5$; the relationship can be
well approximated by linear dependence
\begin{equation}
\ln Q_n(N)=a_n-b_n N\ ,        \label{29}
\end{equation}
where the slope $b_n$ is shown in the inset. Clearly, $b_n$
depends linearly on $n$
\begin{equation}
b_n=\lambda\,n\ ,\quad\quad \lambda=5.35\times 10^{-4}\ .
\label{30}
\end{equation}
We can combine these two equations to write
\begin{equation}
{d\over dN}\ln Q_n(N)=-\lambda\,n\ ,\qquad n\leq 5,    \label{31}
\end{equation}
or
\begin{equation}
Q_n(N)=Q_n(N_0)\,{\rm e}^{-\lambda\,n(N-N_0)}\ .    \label{32}
\end{equation}
Since Eq.\ (\ref{28}) implies $Q_n(N)=\left<x^n\right>_N$, Eq.\
(\ref{32}) is therefore also
\begin{equation}
\langle x^n\rangle_N=\langle x^n\rangle _{N_0}\,{\rm
e}^{-\lambda\,n (N-N_0)}\ .      \label{33}
\end{equation}
In particular, for $n=1, N_0=2$, and $N=350$, we have
\begin{equation}
\langle x\rangle_{350}/\langle x\rangle_2=0.83\ .  \label{34}
\end{equation}
Thus even at the maximum separation between $N$ and $N_0$, the
average $p_T$ of the quarks loses only 17\%. This is because even
severe suppression of high-$x$ quarks cannot change significantly
the average $\left<x\right>$, which is dominated by the low-$x$
behavior. However, the same cannot be said about the higher
moments.

What we have found above are properties of the quarks before
hadronization. Unfortunately, they cannot be checked directly by
experiments. For testable predictions we now go to the study of
proton and kaon formation, whose spectra can be measured
experimentally.
\begin{figure}[tbph]
\includegraphics[width=0.45\textwidth]{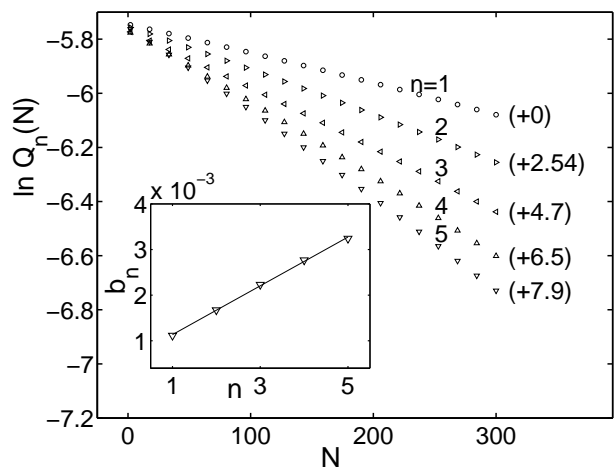}
\caption{Centrality dependence of the moments, $Q_n(N)$,
whose log values are raised by the quantities in the parentheses.
The inset shows the slopes $b_n$, the line being a linear fit.}
\end{figure}

\section{Scaling Distribution of Proton}

The quarks considered in the preceding section can not only
combine with antiquarks to form pions, but also combine with other
quarks to form protons. The formulation of the problem is
discussed in I, where centrality variation is not considered.
Now we allow $N$ to vary, starting from the new quark distribution
$F_q(z)$, whose scaling behavior includes the implicit
dependence on centrality. As before, the treatment of proton
production is not reliable at low $z$, where $p_T$ of the proton is
low enough to make the mass effect important. Our calculation that is
scale invariant cannot take into account the mass-dependent effects.

The proton distribution arising from the recombination of $uud$
quarks is given by \cite{1}
\begin{equation}
{dN_p\over z dz}=\int dz_1dz_2dz_3\,z_1z_2z_3 F_{uud}(z_1,z_2,z_3)
{R}_p (z_1,z_2,z_3,z)\ ,         \label{35}
\end{equation}
where
\begin{equation}
F_{uud}(z_1,z_2,z_3)=F_u(z_1)F_u(z_2)F_d(z_3)\ ,     \label{36}
\end{equation}
and
\begin{equation}
{R}_p(z_1,z_2,z_3,z)=0.057\,z^{-2}G_p(\xi_1,\xi_2,\xi_3)\ .
   \label{37}
\end{equation}
$G_p(\xi_1,\xi_2,\xi_3)$ is the valon distribution in a proton,
expressed in terms of the valon momentum fractions $\xi_i=z_i/z$
\cite{17}. It is determined from the parton distributions that fit
the deep inelastic scattering data, and is
\begin{equation}
G_p(\xi_1,\xi_2,\xi_3)=g\,(\xi_1\xi_2)^\alpha\,\xi_3^\beta\,
\delta(\xi_1+\xi_2+\xi_3-1)\ ,     \label{38}
\end{equation}
where
\begin{equation}
\alpha=1.755,\qquad\qquad \beta=1.05\ ,    \label{39}
\end{equation}
\begin{equation}
g=[B(\alpha+1,\beta+1)B(\alpha+1,\alpha+\beta+2)]^{-1}\ .
\label{40}
\end{equation}
The recombination function for the proton is more complicated than
that for the pion because the proton is not as tightly bound as
the pion in terms of the constituent quarks masses, but the
procedures for the determination of the recombination functions
are similar.

Equation (\ref{35}) is derived in I for $N=N_{\rm max}$. For
$N<N_{\rm max}$ the new scaling function for $F_q$ in Eq.\
(\ref{15}) is to be used for the quark distributions $F_u$ and
$F_d$ in Eq.\ (\ref{36}). The resulting integral is to be
identified with the new proton scaling distribution $\Phi_p(z)$,
as in the pion case. We show in Fig.\ 11 the result on
$\Phi_p(z)$, for which only the $z>3$ part is plotted, since it is
unreliable for $z<3$ due to the neglect of the mass effect.
\begin{figure}[tbph]
\includegraphics[width=0.45\textwidth]{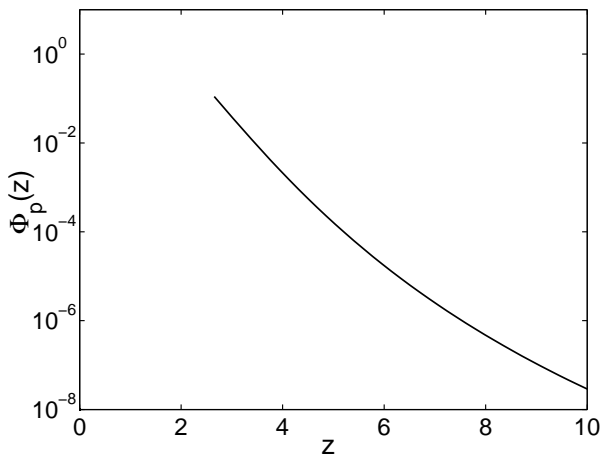}
\caption{ Scaling distribution for proton.}
\end{figure}

The relationship between the calculated $\Phi_p(z)$ and the $p_T$
distribution $dN_p/p_Tdp_T (s,N)$ is
\begin{equation}
\Phi_p(z)=A_p(N)K^2_p(s,N){dN_p\over p_Tdp_T}(s,N)\ ,  \label{41}
\end{equation}
where, we repeat, our $dN_p/p_Tdp_T$ corresponds to the
experimental $(2\pi p_T)^{-1}d^2N_p/d\eta dp_T$ [cf. Eq.\
(\ref{1})]. We have added a subscript $p$ to each function in Eq.\
(\ref{41}) to  emphasize their reference to proton; indeed, we should
similarly add a subscript $\pi$ to the corresponding functions in Eq.\
(\ref{1}) to clarify their differences, as we shall do below when we
compare
$p$ and $\pi$ production. Whereas Eq.\ (\ref{1}) is determined by
the phenomenological analysis of the $\pi^0$ data in II, there
are no similar data on the $p$ spectra to confirm Eq.\ (\ref{41}).
However, on theoretical grounds we expect that relationship to
exist for the following reasons. First, since the quark
distribution is scale invariant, the recombination model implies
that there is a scaling distribution $\Phi_p(z)$ for the proton,
as we have calculated. That gives the LHS of Eq.\ (\ref{41}). On
the RHS we expect $K_p(s,N)=K_\pi(s,N)=K_q(s,N)$, since the same
scaling variable $z$ has been used for pion, quark and proton.
Without that universal variable $z$ the recombination model cannot
be formulated in the form of Eqs. (\ref{11}) and (\ref{35}).
Finally, we conjecture that
\begin{equation}
A_p(N)=A_\pi^{3/2}(N)           \label{42}
\end{equation}
on the  grounds of internal consistency, since the comparison
among Eqs.\ (\ref{1}), (\ref{11}) and (\ref{13}) suggests that
$F_q$ implicitly absorbs an $A_\pi^{1/2}$ factor. When that factor
is applied to Eqs.\ (\ref{35}), (\ref{36}) and (\ref{41}), we
expect Eq.\ (\ref{42}) to follow. This conjecture can be
independently checked when the centrality dependence of the proton
distribution at high $p_T$ becomes available and the existence of
the scaling $\Phi_p(z)$ can then be examined directly from the
data, as is done in II. If the conjecture is verified, then the
data provide empirical evidence for proton being the hadronization
product of three quarks, rather than other mechnisms such as gluon
junction \cite{4, 19}.

One way to check Eq.\ (\ref{42}) is to calculate the $p/\pi$
ratio, $R_{p/\pi}$, at different centrality bins, where
\begin{equation}
R_{p/\pi}(p_T, N)=\left.{d N_p\over p_Tdp_T}(N)\right/{dN_\pi\over
p_Tdp_T}(N)\ .         \label{43}
\end{equation}
In I this ratio has been calculated for $N=N_{\rm max}$. Using
Eqs.\ (\ref{41}) and (\ref{42}) we now can calculate it for
$N<N_{\rm max}$. There is no data in the $p_T$ range where our
prediction is reliable. However, PHENIX does have preliminary data
on $R_{p/\pi}$ for $p_T<3.7$ GeV/c for two centrality bins, 0-5\%
and 60-91\% \cite{14}. They differ by roughly a factor of 3 with
large errors for $p_T\approx 3$ GeV/c. Assuming that the ratio of
ratios is likely to remain the same at higher $p_T$, we can
calculate it for comparison, with the definition
\begin{equation}
r(p_T)=R_{p/\pi}(p_T, N=350)/R_{p/\pi}(p_T, N=15)\ ,   \label{44}
\end{equation}
where $N=15$ is taken to correspond to 60-91\% centrality. The
result is shown in Fig.\ 12 for $\sqrt s=200$ GeV. While the tendency
of
$r(p_T)$ to increase at low $p_T$ is disturbing, the level of
$r(p_T)\approx 3$ for $p_T>5$ GeV/c is in rough agreement with the
data at
$p_T\approx 3$ GeV/c. If $A_p(N)$ were the same as $A_\pi(N)$,
then $r(p_T)$ would be much lower by a factor of 8.4 and can be
ruled out even by the preliminary data \cite{14}. Thus, until
sufficient data become available to test directly the scaling
formula (\ref{41}) for protons, we shall use Eq.\ (\ref{42}) for
the normalization factor.
\begin{figure}[tbph]
\includegraphics[width=0.45\textwidth]{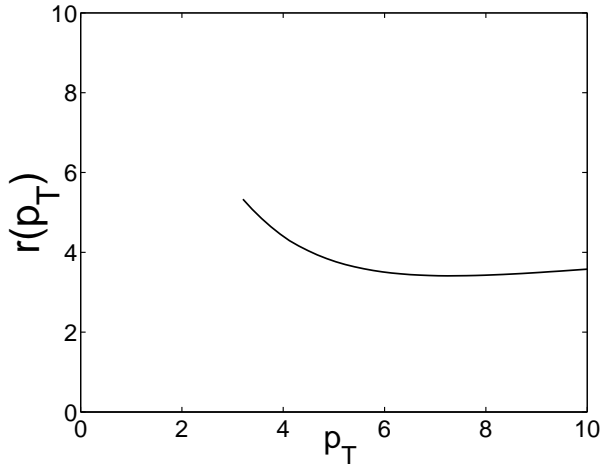}
\caption{The ratio of the $p/\pi$ ratio at $N=350$ to the
same at $N=15$ for $\sqrt s=200$ GeV.}
\end{figure}

On the basis of Eq.\ (\ref{41}) we now can calculate the $p_T$
distributions, $dN_p/p_Tdp_T$, that are experimentally measurable.
They are shown in Fig.\ 13 for $N=30, 150,$ and 350 for $\sqrt s=200$
GeV. It is interesting to note that the distributions for $N=150$ and
350 differ only slightly in the log scale, due undoubtedly to the near
cancellation of the two opposing properties: the increase of the
number of hard collisions at higher $N$ and the suppression of
high $p_T$ protons in larger dense medium.
\begin{figure}[tbph]
\includegraphics[width=0.45\textwidth]{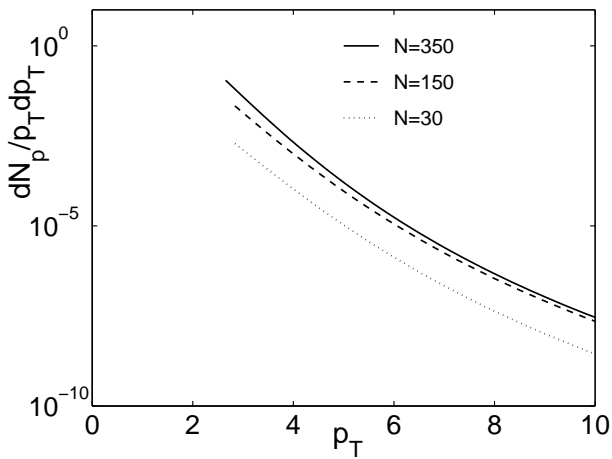}
\caption{The $p_T$ distribution for proton at three values
of $N$ for $\sqrt s=200$ GeV.}
\end{figure}

\section{Scaling Distribution of Kaons}

For the production of kaons in the recombination model we need two
inputs: the strange quark distribution and the recombination
function for a $K$ meson. The former has been considered in our
study of the strangeness enhancement problem in heavy-ion
collisions \cite{18}. The latter is given in Ref.\ \cite{13}.

It is known that the number of $\Lambda$ and $\Sigma$ hyperons
produced in heavy-ion collisions is enhanced by more than a
factor of 2. Since the enhancement is due mainly to gluon
conversion and Pauli blocking, the relevant question in the context
of the parton model is what the percentage of the gluon conversion
into the strange quarks is. The amazing answer found in Ref.\
\cite{18} is that it is only 8\%. Due to the large number of
gluons produced in heavy-ion collisions that is enough to raise
the strange quark density by a factor of 2.1 to 2.3 from the
intrinsic level in a free proton, depending on collision energy.

To be precise, let us use the notation where the symbols $\ell$
and $s$ denote the total number of light and strange quarks,
respectively, $\ell$ being $u+d$, in contrast to $q$ being $u$ or
$d$. Let $\ell_v, \ell_s, s_s$ and $g$ denote the numbers of
valence quarks, light sea quarks, strange quarks and gluons,
respectively, before the gluons are converted to $\ell\bar\ell$
and $s\bar s$ for recombination to form hadrons. Since only the ratios
of these numbers will be relevant below, they will be given modulo
a common multiplicative factor that need not be specified. The
numbers for RHIC at 130 GeV are \cite{18}
\begin{eqnarray}
\ell_v&=&0.30\ , \quad\quad \ell_s=0.37\ ,\\ \nonumber
s_s&=&0.18\ , \quad\quad g=2.46\ .   \label{45}
\end{eqnarray}
It is also found that the fraction of gluon conversion to the
strange sector is $\gamma=0.08$, i.e.,
\begin{equation}
s_c=\gamma g\ ,\quad\quad \ell_c=(1-\gamma)g\ ,    \label{46}
\end{equation}
where the subscript $c$ denotes converted quarks. The net strange
to light quark ratio after conversion is then
\begin{equation}
{s\over \ell}={s_s+s_c\over \ell_v+\ell_s+\ell_c}=0.128\ .
         \label{47}
\end{equation}
Setting $u=d$ for simplicity, we have $\ell=2q$ and
\begin{equation}
s/q=0.256\ .            \label{48}
\end{equation}
Since $s_c=\gamma g=0.197,$ the strange quark multiplicity
originally at $s_s$ is more than doubled by gluon conversion.

The above consideration is at the quark level. How that
translates to hadron abundance must take into account hyperon
production in addition to kaon production, since the effects of
associated production cannot be ignored. The problem of
partitioning the total strange quark numbers into various channels
of strange hadrons competing for those quarks has been treated in
Ref.\ \cite{18}. It is found there that the fraction $\kappa$ of
$s$ quark forming $\bar K$ and $\bar\kappa$ of $\bar s$ antiquark
forming $K$ can be deduced from the data. That is, defining
\begin{equation}
\bar{K}=\kappa s\ ,\quad\quad K=\bar{\kappa}\bar{s}\ ,   \label{49}
\end{equation}
where $\bar K=K^-+\bar K^0$ and $K=K^++K^0$ denote the numbers of $K$
mesons of various types, one has at 130 GeV collision energy
\begin{equation}
\kappa=0.628\ ,\quad\quad \bar{\kappa}=0.713\ .    \label{50}
\end{equation}
For the purpose of calculating $K^+/\pi^+$ ratio, let us use
$\bar s'$ to denote the number of $\bar s$ quarks to recombine
with $u$ quark to form $K^+$, and we have
\begin{equation}
\bar{s}^\prime={0.256\over 2}\bar{\kappa} q=0.091q\ .  \label{51}
\end{equation}
We shall assume that the $z$ distribution of the $\bar s'$ quark
is the same, apart from normalization, as that of the $u$ quark,
so we get
\begin{equation}
F_{\bar s'}(z)=0.091F_q(z)\ .     \label{52}
\end{equation}

In the recombination model we expect the produced $K^+$ to have
also a scaling distribution
\begin{equation}
\Phi_K(z)=\int dz_1dz_2\,z_1z_2 F_q(z_1)F_{\bar s'}(z_2){
R}_K(z_1,z_2,z)\ .    \label{53}
\end{equation}
The recombination function for the $K$ meson is similar to that of
the pion given in I
\begin{equation}
R_K(z_1,z_2,z)=R_K^0\,z^{-2}\,G_K\left({z_1\over z},{z_2\over
z}\right),     \label{54}
\end{equation}
where $G_K(\xi_1,\xi_2)$ is the valon distribution in the $K$ meson
\cite{13}
\begin{equation}
G_K(\xi_1,\xi_2)=g_K\,\xi_1^a\xi_2^b\,\delta(\xi_1+\xi_2-1)
    \label{55}
\end{equation}
with $g_K=B(a+1,b+1)^{-1}$. The parameters $a$ and $b$ are
determined from the analysis of $Kp$ collisions and are found to
be $a=1$ and $b=2$ (see the second paper in Ref.\ \cite{13}).
Since $K^+$ is the only state in the pseudoscalar octet that has
$u\bar s$ content, we have $R_K^0=1$ and thus
\begin{equation}
R_K(z_1,z_2,z)={1\over B(2,3)}{z_1z_2^2\over
z^5}\,\delta\left({z_1\over z}+{z_2\over z}-1\right).   \label{56}
\end{equation}
We now can calculate $\Phi_K(z)$ using Eqs.\ (\ref{15}),
(\ref{52}), (\ref{53}), and (\ref{56}). Since for $\Phi_K(z)$ we
have, as in Eq.\ (\ref{41}),
\begin{equation}
\Phi_K(z)=A_K(N)K^2_K(s, N){dN_K\over p_Tdp_T}(s, N)\ ,
\label{57}
\end{equation}
where $A_K=A_\pi$ and $K_K=K_\pi$, we obtain for the $K^+/\pi^+$
ratio
\begin{equation}
R_{K/\pi}(p_T, N)=\left.{dN_K\over p_Tdp_T}\right/{dN_\pi\over
p_Tdp_T}={\Phi_K(z(p_T, N))\over \Phi_\pi(z(p_T, N))}\ ,
  \label{58}
\end{equation}
where $\pi^+$ is taken to be the same as $\pi^0$.

In Fig.\ 14 we show $R_{K/\pi}(p_T)$ for $N=350$ and 200 and $\sqrt
s=130$ GeV. Since the determination of $\kappa$ and $\bar\kappa$ in
Eq.\ (\ref{50}) is by use of the data on particle ratios, which are
not reliable for non-central collisions, we have no confidence in the
strangeness enhancement factor deduced when $N$ is low. For
$N>200$ Fig.\ 14 shows that the $K/\pi$ ratio is not sensitive to
centrality, and only mildly dependent on $p_T$. This result awaits
direct check by experimental data.
\begin{figure}[tbph]
\includegraphics[width=0.45\textwidth]{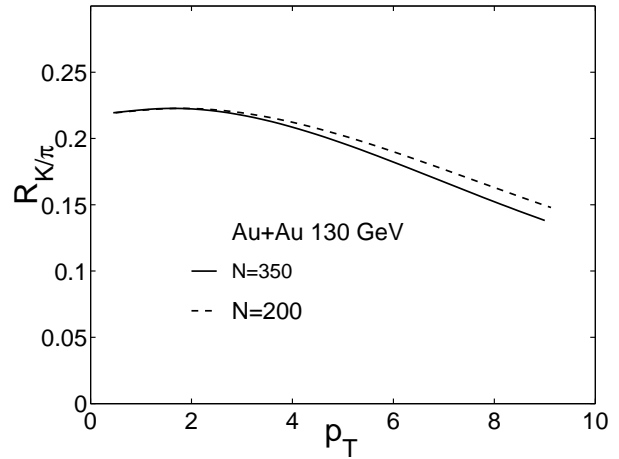}
\caption{The $K/\pi$ ratio at two values of $N$ for $\sqrt
s=130$ GeV.}
\end{figure}

\section{Conclusion}

From the scaling distribution $\Phi(z)$ obtained from the data on
$\pi^0$ production, we have derived many quantities by use of the
recombination model. They are the distributions of quarks,
protons, and kaons, and their respective dependences on
centrality. All that is made possible by the discovery  of the
universal function $\Phi(z)$ valid for all centrality and
collision energy. The scaling variable $z$ that unites the $p_T$
dependences for all $N$ and $\sqrt s$ quantifies the difficulty of
producing transverse motion, and for that interpretation it can be
termed {\it transversity}. A particle produced at a particular
$p_T$ at high $\sqrt s$ has a lower transversity than that of the
same particle produced at the same $p_T$ but at a lower $\sqrt s$,
because it is easier in the former case that involves a smaller
momentum fraction. Similarly, it is easier to produce a particle at
a given high $p_T$ when $N$ is low than to do the same at a higher
$N$ because there is less degradation of the transverse momentum at
lower $N$. Thus the former particle has lower transversity than the
latter. With this way of viewing transverse motion we see that it is
the transversity of the produced particles that has the universal
property at whatever centrality and collision energy.

The quantitative results that we have obtained have their
limitations due to the assumptions that we have made. For example,
for the proton $z$ distribution we have assumed that $F_{\bar
q}(z)/F_q(z)$ is a constant, which has some phenomenological
support in that the observed $\bar p/p$ ratio is roughly constant
in $p_T$; however, that ratio fails to maintain constancy at very
peripheral collisions. Thus our result is not likely to be valid
when $N$ is very low. Similarly, whether $F_{\bar s'}(z)/F_q(z)$
is a constant and over what range of $N$, if it is, are not
known. The centrality dependence of the strangeness fraction of
gluon conversion is also unknown. Our result on $K/\pi$ ratio can
therefore only be regarded as preliminary, pending experimental
guidance to improve our simplifying assumptions. Despite these
uncertainties for non-central collisions, the recombination model
has enabled us to calculate the scaling behavior of the quark,
proton, and kaon distributions, the latter two of which are
subject to direct experimental test.

For non-central collisions we have only calculated the
$p_T$ distribution, averaged over the azimuthal angle $\phi$.
Clearly, the dependence on $\phi$ is important as it contains
dynamical information. It will be very interesting to investigate
whether centrality scaling persists in restricted $\phi$ bins, and
if it does, how the scaling curves depend on $\phi$. On the basis
of the universality in transversity distribution, we expect that
such $\phi$-dependent scaling curves can be put into an overall
$\phi$-independent scaling curve upon rescaling. At the price of
lower statistics this can be checked by appropriate analysis of
the $\pi^0$ production data.

Ultimately, the important issue to focus on is the implication of
the existence of the scaling behavior on the possible formation of
quark-gluon plasma. At this stage of our understanding a
conservative statement that can be made is that the discovery of
scaling violation might provide a strong hint for a drastic change
of dynamics, possibly associated with a phase transition. Without
waiting for LHC to enlighten us with that possibility, a more
urgent issue to settle is why, if the quark-gluon plasma has been
created already at existing collision energies, the change of
the nature of the dense medium does not affect the scaling behavior
that we have found to be universal between 17 and 200 GeV. Either
the scaling behavior is insensitive to the change, or the change
has already occurred at collision energy less than 17 GeV. If both
of these alternatives are incorrect, then the only way out is that
quark-gluon plasma has not yet been created at RHIC. These are the
unintended, but remarkable, consequences of the scaling behavior.
It is thus paramount to understand whether a phase transition can
lead to a violation of the scaling behavior found here.

\section*{Acknowledgment}
We are grateful to W.\ A.\ Zajc and D.\ d'Enterria for their
very helpful comments. This work was supported, in part,  by the U.\
S.\ Department of Energy under Grant No. DE-FG03-96ER40972.

\end{document}